# Comparison of Two Detector Magnetic Systems for the Future Circular Hadron-Hadron Collider

Vyacheslav Klyukhin [1,2,*], Austin Ball [2], Christophe Paul Berriaud [3], Benoit Curé [2], Alexey Dudarev [2], Andrea Gaddi [2], Hubert Gerwig [2], Alain Hervé [4], Matthias Mentink [2], Werner Riegler [2], Udo Wagner [2] and Herman Ten Kate [2]

[1] Skobeltsyn Institute of Nuclear Physics, Lomonosov Moscow State University, RU-119992 Moscow, Russia
[2] European Organization for Nuclear Research (CERN), CH-1211 Geneva 23, Switzerland; austin.ball@cern.ch (A.B.); benoit.cure@cern.ch (B.C.); alexey.dudarev@cern.ch (A.D.); andrea.gaddi@cern.ch (A.G.); hubert.gerwig@cern.ch (H.G.); matthias.mentink@cern.ch (M.M.); werner.riegler@cern.ch (W.R.); udo.wagner@cern.ch (U.W.); herman.tenkate@cern.ch (H.T.K.)
[3] CEA Irfu, 91191 Saclay, France; christophe.berriaud@cea.fr
[4] Department of Physics, University of Wisconsin, Madison, WI 53706, USA; alain.herve@cern.ch
* Correspondence: vyacheslav.klyukhin@cern.ch

**Featured Application:** This work describes a detailed study of two possible options for the magnetic system of a Future Circular hadron-hadron Collider detector.

**Abstract:** The conceptual design study of a Future Circular hadron-hadron Collider (FCC-hh) to be constructed at CERN with a center-of-mass energy of the order of 100 TeV requires superconducting magnetic systems with a central magnetic flux density of an order of 4 T for the experimental detectors. The developed concept of the FCC-hh detector involves the use of an iron-free magnetic system consisting of three superconducting solenoids. A superconducting magnet with a minimal steel yoke is proposed as an alternative to the baseline iron-free design. In this study, both magnetic system options for the FCC-hh detector are modeled with the same electrical parameters using Cobham's program TOSCA. All the main characteristics of both designs are compared and discussed.

## 1. Introduction

The conceptual design report for the Future Circular Collider (FCC) [1] with a center-of-mass energy of the order of 100 TeV, assumed to be constructed at CERN in a new tunnel of 80–100 km circumference, was presented in 2019 [2]. As part of this study, the concept of an FCC-hh detector for hadron-hadron physics was developed [3]. Apart from the particle sub-detectors, the FCC-hh detector will comprise an iron-free magnetic system consisting of three superconducting solenoids: the main coil with a central magnetic flux density of 4 T and two auxiliary forward solenoid coils with a central magnetic flux density of 3.2 T each [4].

Following the alternative study of the magnetic system for the FCC-hh detector based on the superconducting solenoid with a minimal steel yoke [5], the possibility of using the steel yoke in the same baseline layout of the coils and the particle sub-detectors is considered, and the main parameters of both magnetic systems are compared here. For these comparisons, both the baseline FCC-hh detector magnetic system [4], and the magnetic system with the minimal steel yoke, are calculated with the program TOSCA from Cobham CTS Limited [6].

A part of this study was presented in 2018 at 6th International Conference on Superconductivity and Magnetism—ISCM2018 in Antalya, Turkey. At the time, the minimal steel yoke option included exact shapes of the radiation protection shields around the forward coils in correspondence with the baseline magnetic system. In the baseline option, these shields are assumed to be constructed from a nonmagnetic material. In the minimal steel yoke option, the shields are the parts of the steel flux return yoke. The original shapes of these shields include the large cones propagated inside the main coil, that in the minimal steel yoke design creates the enormous magnetic flux density inside the cones and, as a result, the large axial electromagnetic forces towards the center of the main coil. In this study, these cones are assumed to be from a nonmagnetic material to reduce the axial forces and the perturbation of the magnetic flux density in the endcap parts of the yoke.

To study the quality of the magnetic field inside the tracking system volume, the magnetic field double integral method [7] is used instead of an investigation of the magnetic field bending power for the charged particles.

The paper is organized in a way as follows: Section 2 contains a description of both options of the FCC-hh detector magnetic systems; Section 3 includes general comparisons of both options and a detailed investigation of the magnetic field quality in terms of the magnetic field double integral method; Section 4 represents a small discussion of the obtained results; and finally, conclusions are presented in Section 5.

## 2. Modeling the Magnetic Systems

### 2.1. Reference Geometry of the FCC-hh Detector

A reference geometry for the FCC-hh detector with the baseline magnetic system is shown in Figure 1. The detector is designed to register particles generated in collisions of hadron beams at a center-of-mass energy of 100 TeV and assumed to have a diameter of 20 m and a length of 50 m. This is comparable to the dimensions of the ATLAS [8] and CMS [9] detectors at the Large Hadron Collider [10] at CERN. The central part of the detector houses the tracking system, and electromagnetic and hadron calorimetry inside a 4 T superconducting solenoid with a free bore diameter of 10 m. The forward parts of the detector are displaced from the beam interaction point along the beam axis and two forward solenoid coils with an inner bore of 5 m provide the required bending power for the charged particles escaping the interaction point.

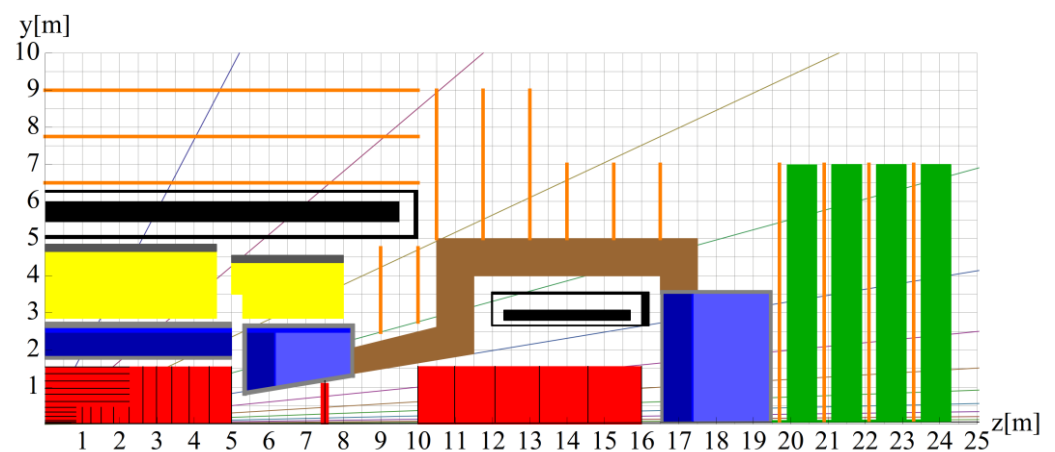

**Figure 1.** One quarter of the reference geometry for the FCC-hh detector with the baseline magnetic system. In black, the quarter of the main superconducting coil in the vacuum cryostat and a half of the forward coil in its own cryostat are shown. In brown, the radiation protection conical-cylinder shield made of nonmagnetic material is displayed. In red, the charged particle tracker cylinders are shown. The orange lines represent the locations of the muon detection chambers. Other colors describe electromagnetic and hadronic calorimeters, and nonmagnetic disks of the forward muon system [3]. In this study, regions of these nonmagnetic materials are considered to have free-space permeability.

Increasing the particle beam colliding energy to 100 TeV leads to an increase in transverse momenta of the charged particles to be detected in the tracking system up to at least 10 TeV/*c*. The tracker is specified to provide better than 20% momentum resolution for these particles and better than 0.5% momentum resolution at the multiple scattering limit [3]. To achieve this goal, the inner tracker, consisting of three cylinders, has a 1.55 m radius and a length of 32 m.

The scheme of the FCC-hh detector shown in Figure 1 includes the parts of interest for this study: the main and forward superconducting coils, the charged particle tracker cylinders, and the conical-cylinder nonmagnetic radiation protection shield.

*2.2. Baseline Magnetic System Model*

The model of the FCC-hh detector baseline magnetic system shown in Figure 2 includes three components: the main superconducting coil of 10.9 m inner diameter and 18.954 m length with a total current of 69.6 MA-turns, and two superconducting forward coils of 5.6 m inner diameter and 3.3997 m length with a total current of 12.6 MA-turns each.

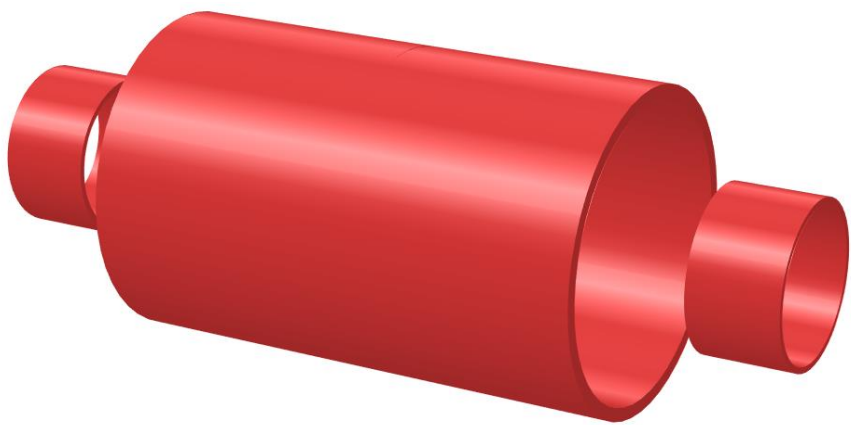

**Figure 2.** Three-dimensional model of the FCC-hh detector baseline magnetic system comprising the main superconducting coil with 10.9 m inner diameter, and two superconducting forward coils with 5.6 m inner diameter. The distance between the coils is 2.823 m each side.

The main coil has about the same central magnetic flux density as the Compact Muon Solenoid (CMS) [9], the largest detector coil in the world, but 1.52 times larger inner diameter and length.

In the model, the radial thickness of the FCC-hh detector main coil is assumed to be 0.5 m, and the radial thickness of each forward coil is 0.23 m. The distance between the main and each forward coil is 2.823 m. Preparing the design of the baseline magnetic system, this distance is chosen as a compromise between the value of the magnetic flux density in the transition region between the coils, and the attractive axial force onto the forward coil that is minimized to 61.8 MN. The compression axial force in the main coil middle plane is 610.7 MN that creates a pressure of 34.1 MPa. The radial pressure is 1.55 MPa in the main coil, and 2.66 MPa in each forward coil. The stored energy of the baseline magnetic system is 13.8 GJ.

*2.3. Minimal Steel Yoke Magnetic System*

To realize the minimal steel yoke conception [5], the main coil is assumed to be surrounded by five barrel wheels of 17.5 m outer diameter and 3.9 m width each, as shown in Figure 3. Each barrel wheel has two 0.75 m thick steel layers with a radial gap of 0.5 m between them.

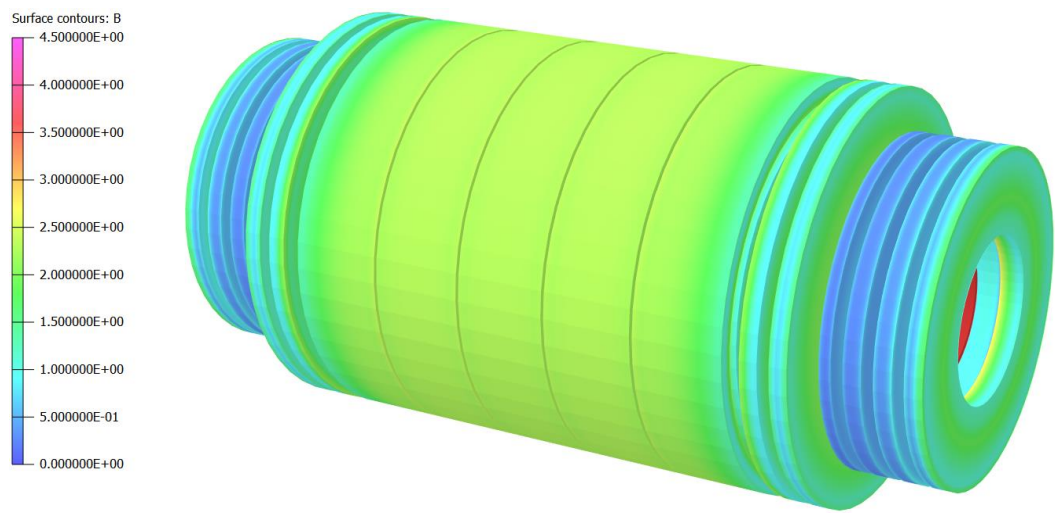

**Figure 3.** Distribution of the magnetic flux density on the minimal steel yoke surface of the FCC-hh detector. The color scale unit is 0.5 T. With the central magnetic flux density of 4.24 T, the minimum and maximum magnetic flux density values on the yoke surface are 0.2037 and 4.0357 T.

At the distance of 0.75 m off the extreme wheels, two endcap disks of 17.5 m outer diameter and 0.75 m thickness are located at each barrel end. Four smaller disks of 14 m outer diameter at each yoke end follow these disks. The thicknesses of the first small disks on both sides of the yoke are 0.5 m, while other disks are 0.75 m in thickness.

The presence of 0.5 m air gaps between all the endcap disks allow room for installation of the muon detection chambers located in the same positions as in the baseline design of the FCC-hh detector [3] shown in Figure 1. As displayed in Figure 4, the inner parts of the endcap disks rely on the cylindrical radiation protection shields of 1 m radial thickness assumed to be made of a carbon steel. These shields have a length of 7 m and an outer diameter of 10 m each.

In contrast, in the baseline design of the magnetic system shown in Figure 1, these shields are assumed to be made of nonmagnetic material. The total length of the minimal steel yoke is 35 m.

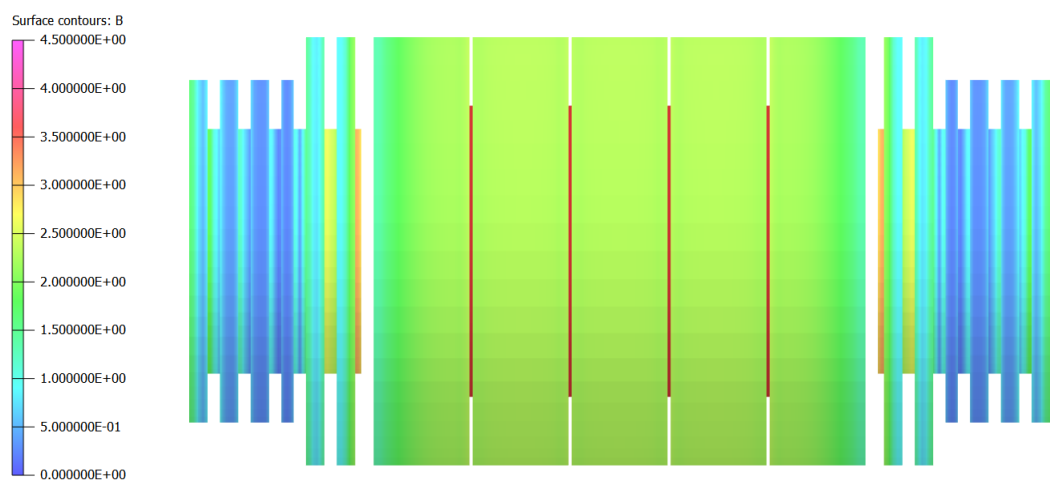

**Figure 4.** The main superconducting coil with an outer diameter of 11.9, the five barrel wheels measuring 3.9 m in width, the two cylindrical radiation protection shields of 10 m outer diameter and 7 m long around each forward coil, the four endcap disks of 17.5 m diameter each, and the eight endcap disks of 14 m diameter each are shown. The main solenoid coil is visible between the barrel wheels in the air gaps of 0.125 m each. The length of the barrel part around the main coil is 20 m; the total length of the yoke is 35 m. The color scale of the magnetic flux density on the yoke surface is the same as in Figure 3.

The main superconducting coil and the forward solenoid coils have the same dimensions and the same total currents as in the baseline magnetic system design.

## 3. Comparison of the Magnetic Systems

### 3.1. General Comparisons

In Figures 5 and 6, the magnetic flux density distribution is displayed in a vertical $YZ$ plane of the baseline and minimal steel yoke magnetic systems, respectively. The magnetic flux density in the center of the main coil is 4 T in the baseline and 4.24 T in the minimal steel yoke designs, accordingly.

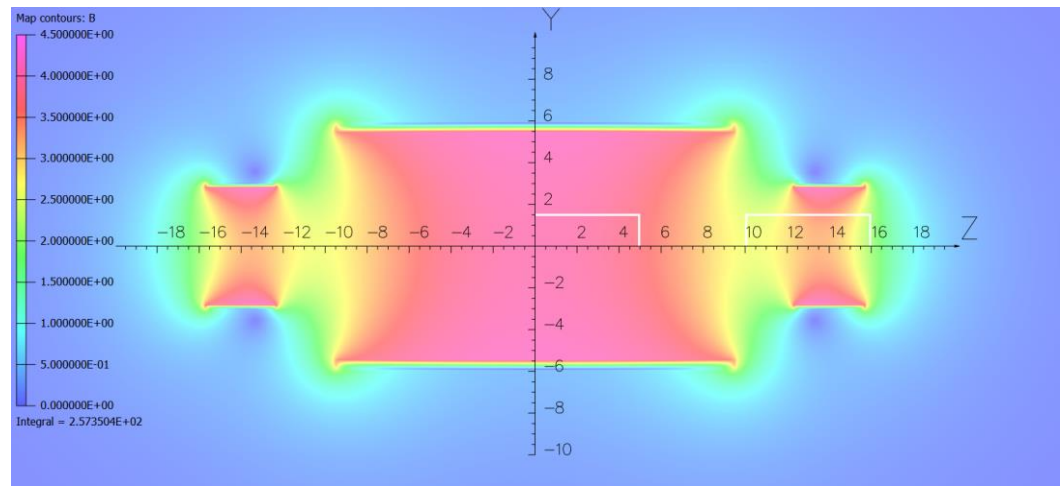

**Figure 5.** Magnetic flux density distribution in a vertical $YZ$ plane of the FCC-hh detector baseline magnetic system. The color magnetic field map plotted with the cell size of 0.05 m has a width of 50 m and height of 23 m. The color scale unit is 0.5 T. The minimum and maximum magnetic flux density values are 0.0142 and 4.1595 T. The quarter of the tracking system used in this analysis is drawn with the white rectangles.

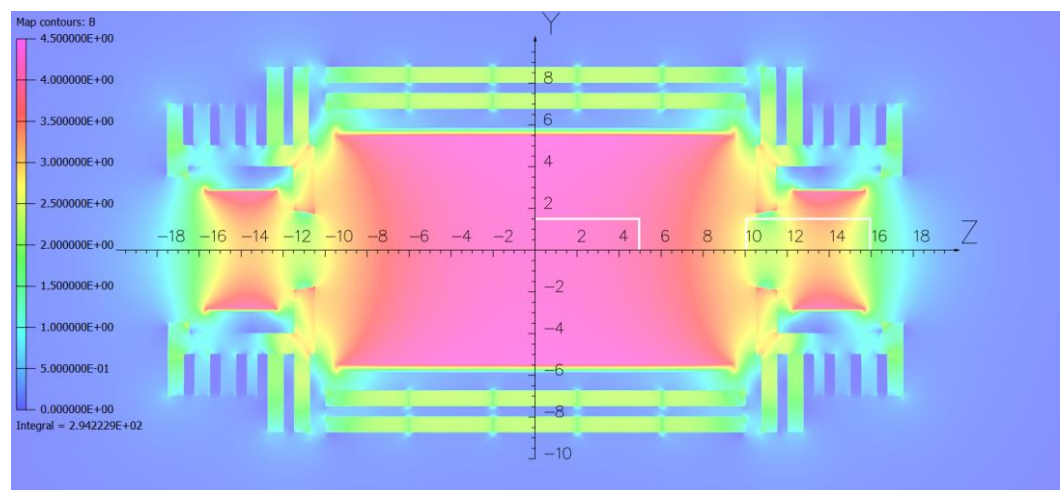

**Figure 6.** Magnetic flux density distribution in a vertical plane of the FCC-hh detector minimal steel yoke magnetic system. The color magnetic field map plotted with the cell size of 0.05 m has a width of 50 m and height of 23 m. The color scale unit is 0.5 T. The minimum and maximum magnetic flux density values are 0.0002 and 4.3623 T. The maximum magnetic flux density in the barrel wheel layers is 2.3 T. The magnetic flux density in the first endcap disks at the radius of 6 m is 2.4 T. The quarter of the tracking system used in the present analysis is drawn with the white rectangles.

The larger central magnetic flux density value in the minimal steel yoke option reflects the contribution of the yoke to the coil's inner magnetic field.

In both figures, the color scales for the magnetic flux density are the same: from 0 to 4.5 T. The same color in both figures corresponds to the same magnetic flux density value. The magnetic flux distributions in these figures are somewhat different, and some areas of the systems are described with slightly different colors.

Comparison of the magnetic flux distribution in Figures 5 and 6 displays that the magnetic flux inside the coils is very similar for both options except in the transition areas between the main and forward coils, where the magnetic flux density in the minimal steel yoke option is lower than in the baseline design by 9.3%. The presence of an inner steel disk with a thickness of 1 m and a bore diameter of 3.47 m causes this effect.

As displayed in Figure 7, the 6% increase in the central magnetic flux density, caused by the steel yoke contribution, is compensated by a decrease in the magnetic flux density at the axes of the forward coils. At distances of ±13.53 m from the main coil center along the coil axes, the magnetic flux density is 3.2 T in the baseline design and 3.04 T (5% lower) in the minimal steel yoke magnetic system.

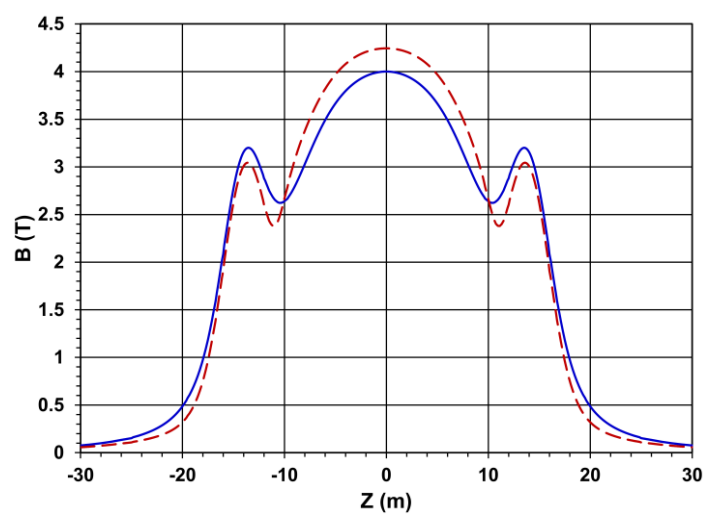

**Figure 7.** Magnetic flux density variation along the coil axes in the baseline (smooth curve) and minimal steel yoke (dashed line) magnetic systems.

Figure 8 presents the stray magnetic flux density variation vs. radius in the middle plane of the main coil, and vs. distance from the main coil center along the coil axes.

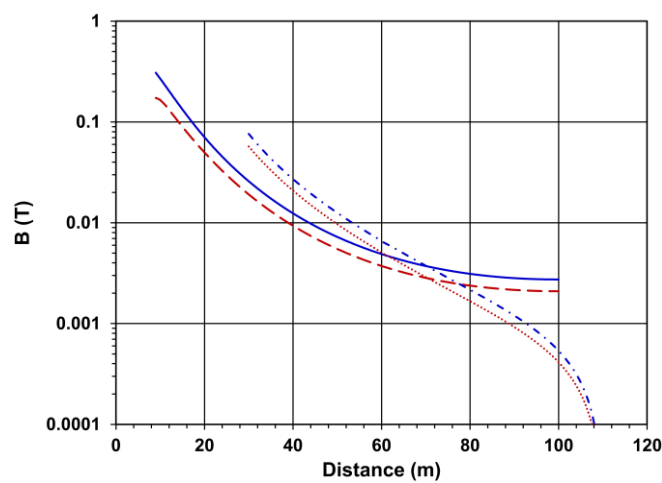

**Figure 8.** Magnetic flux density out of the coil in the main coil central plane vs. radius (smooth and dashed lines) as well as along the coil axes vs. distance from the main coil center (dash-dotted and small-dotted lines). Smooth and dash-dotted lines correspond to the baseline magnetic system. Dashed and small-dotted lines correspond to the minimal steel yoke magnetic option.

The stray magnetic flux density drops to the safe level of 5 mT at a radius of 59.4 m (52.2 m) from the coil axis in the baseline (minimal steel yoke) design. The stray field drops to 5 mT at 64.9 m (60.2 m) from the main coil center along the coil axes in the baseline (minimal steel yoke) design.

In the minimal steel yoke magnetic systems, the compression axial force onto the central barrel wheel is 236.9 MN from each side including the attractive axial force onto the endcap assembly of 164 MN. These values are 2.92 and 2.42 times larger than the corresponding axial forces in the CMS magnet yoke at the 3.81 T central field [11].

The axial force on each forward coil in the minimal steel yoke magnetic system is 62.4 MN against of 61.8 MN in the baseline design. The compression axial force in the main coil middle plane is 529.8 MN (against of 610.7 MN) that creates a pressure of 29.6 MPa (against of 34.1 MPa). The radial pressure in the main coil is 2.5 MPa (against of 1.55 MPa), and in each forward coil is of 2.23 MPa (against of 2.66 MPa). The stored energy is 14.6 GJ that is 5.8% larger than in the baseline design, and 6.4 times larger than in the CMS magnet [9].

*3.2. Magnetic Field Double Integral Comparisons*

To quantify the quality of the magnetic field inside the inner tracker, the magnetic field double integrals method is applied. This method is fully described in Ref. [7] and was used to characterize the quality of the magnetic field in different options [12,13] of the magnetic system for the FCC-hh detector [3].

As shown in Figure 1, the considered inner tracker of the detector consists of three cylinders with a radius of 1.55 m each sitting on the coil axes. The central cylinder has a length of 10 m and is located inside the main superconducting coil. The other two cylinders are 6 m long and are located inside the forward coils at the distances of 5 m from the central tracker cylinder. In the following study, only one quarter of the tracking system in the vertical plane is considered. This quarter is shown in Figures 5 and 6 as white rectangles of 1.55 m radius propagated along the *Y* axis. A half length of 16 m is directed along the *Z* axes in the same figures.

In this region from zero to $R_{max}$ = 1.55 m and from zero to $Z_{max}$ = 16 m, the magnetic field double integrals are calculated along the trajectories of the charged particles emitted in a vertical *YZ* plane from the origin of the coordinate system at different polar angles $\theta$ counted from the *Z*-axis. A pseudorapidity value $\eta$, defined as $\eta = -ln[tan(\theta/2)]$, corresponds to each polar angle.

Three specific pseudorapidity values are referred to the corners of the inner tracker volumes in the vertical plane. The first value $\eta_1$ = 1.88753 refers to the corner of the central tracking cylinder at $R$ = 1.55 and $Z_1$ = 5 m. The second value $\eta_2$ = 2.56343 refers to the front corner of the forward tracking cylinder at $R$ = 1.55 and $Z_2$ = 10 m. The third value $\eta_c$ = 3.02982 refers to the corner of the forward tracking cylinder at $R$ = 1.55 and $Z_{max}$ = 16 m. The magnetic field double integrals are calculated in the pseudorapidity range from 0 to 4 with an increment of 0.05. The polar angle $\theta$ corresponding to $\eta$ = 0 is equal to 90°. The polar angle $\theta$ corresponding to $\eta$ = 4 is equal to 2.0986°.

Inside the central tracking cylinder while the pseudorapidity values are smaller than $\eta_1$, the full track length *L* registered in the tracking volume is equal to $R_{max}/sin\theta$. For the values larger than $\eta_1$ but smaller than $\eta_2$, the length *L* is equal to $Z_1/cos\theta$, where $R_{max}$ and $Z_1$ are the radius and half a length of the central inner tracker volume. In the pseudorapidity region from $\eta_2$ to $\eta_c$, the track length *L* is again determined by the ratio $R_{max}/sin\theta$. Finally, for pseudorapidity greater than $\eta_c$, the total track length *L* is equal to $Z_{max}/cos\theta$.

The magnetic field double integral $I_2$ is determined by Equation (1) as follows [7]:

$$I_2 = \int_0^{L\,sin\theta} \int_0^{r/sin\theta} B\, sin\theta_{(d\mathbf{l},\mathbf{B})}\, dl\, dr. \tag{1}$$

Here, $r$ is a running transverse radius and $l$ is a running track length along the track length $L$ during the double integration of a product of the total magnetic flux density $B$ and the track direction. The polar angle $\theta_{(dl, B)}$ represents the longitudinal component of the angle between the track projection to the vertical plane and the magnetic flux density vector, i.e., both the track length and the magnetic flux density vector are considered to lie in the vertical plane [7].

In the constant homogeneous magnetic field, the integral $I_2$ is equal to $B \cdot R_{max}^2/2$ in the pseudorapidity regions $\eta < \eta_1$ and $\eta_2 < \eta < \eta_c$ and drops like $B \cdot Z_{max}^2 \cdot \tan^2\theta/2$ in the pseudorapidity region $\eta > \eta_c$. In the inhomogeneous magnetic field, the double integral $I_2$ degrades by a few percent in comparison with the ideal constant magnetic field.

In the silicon tracker with $N$ equidistant detector planes and a resolution $\sigma$ [m] per layer, the momentum resolution at the large transverse momenta $p_T$ [GeV/c] is dominated by the detector resolution, and the relative transverse momentum precision $\delta$ can be expressed as follows [14,3]:

$$\delta = \frac{dp_T}{p_T} \approx \frac{\sigma p_T}{0.3 I_2} \sqrt{\frac{720}{N+4}}, \qquad (2)$$

where $I_2$ [T·m$^2$] is determined by Equation (1).

In Figure 9, the dependence of the magnetic field double integrals vs. pseudorapidity is presented for four cases:

- $I_{2bh}$—double integral in homogeneous constant magnetic field of 4.000174 T, that corresponds to the central magnetic flux density in the baseline configuration of the magnetic system.
- $I_{2bi}$—double integral in the real inhomogeneous magnetic field of the baseline configuration of the magnetic system.
- $I_{2mh}$—double integral in homogeneous constant magnetic field of 4.243783 T, that corresponds to the central magnetic flux density in the minimal steel yoke configuration of the magnetic system.
- $I_{2mi}$—double integral in the real inhomogeneous magnetic field of the minimal steel yoke configuration of the magnetic system.

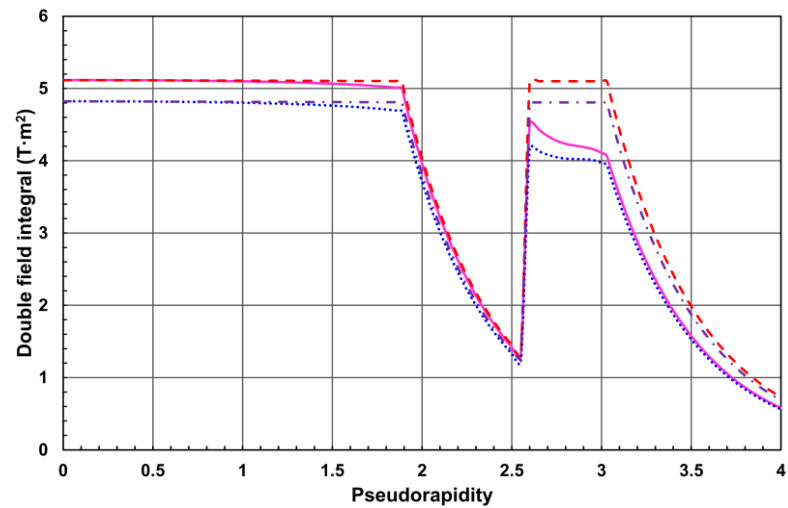

**Figure 9.** Magnetic field double integrals in a vertical plane of the inner tracker of the minimal steel yoke design (smooth—$I_{2mi}$ and dashed—$I_{2mh}$ lines) and of the baseline FCC-hh detector design (small-dotted—$I_{2bi}$ and dash-dotted—$I_{2bh}$ lines) vs. pseudorapidity.

In the pseudorapidity regions $\eta < \eta_1$, and $\eta_2 < \eta < \eta_c$ the double integral $I_{2bh}$ has a constant value of 4.81 T m$^2$, the double integral $I_{2bi}$ drops from 4.82 to 4.69 T m$^2$, and from 4.19 to 3.97 T m$^2$, correspondingly. In the same pseudorapidity regions, the double integral $I_{2mh}$ has a constant value of 5.1 T m², the double integral $I_{2mi}$ drops from 5.12 to 5.0 T m², and from 4.52 to 4.11 T m², accordingly. In the pseudorapidity regions $\eta_1 < \eta < \eta_2$ and $\eta > \eta_c$, all the double integrals drop like ~$tan^2\theta$.

Further comparisons are performed for the variables as follows: a ratio $R_{bh} = I_{2bi}/I_{2bh}$, a ratio $R_{mh} = I_{2mi}/I_{2mh}$, and a ratio $R_{mb} = I_{2mi}/I_{2bi}$, where indexes $h$, and $i$ stand for the homogeneous and inhomogeneous magnetic field, accordingly. The indexes $b$ and $m$ denote the baseline and minimal steel yoke magnetic systems, correspondingly.

As reflected in Figure 9, in the pseudorapidity range from 0 to 2.55 the ratios $R$ stay in the limits as follows: $1.0006 > R_{bh} > 0.9737$, $1.0004 > R_{mh} > 0.9801$, $1.0607 < R_{mb} < 1.0678$.

In the pseudorapidity range from 2.6 to 4, the ratios $R$ stay in the limits as follows: $0.8706 > R_{bh} > 0.8169$, $0.8869 > R_{mh} > 0.7949$, $1.0808 > R_{mb} > 1.0324$.

These results show that both integrals $I_{2hi}$ and $I_{2mi}$ degrade with pseudorapidity comparing with the ideal magnetic field double integrals $I_{2bh}$ and $I_{2mh}$, correspondingly, but the integral $I_{2mi}$ is always larger than integral $I_{2bi}$ by 3.2–8.1%. According to Ref. [7], larger magnetic field double integral produces larger track sagitta on the same track length $L$ with the same transverse momentum $p_T$.

From Equation (2), the relative momentum precision for a real versus ideal solenoid field is given by a ratio $R = \delta_h/\delta_i$, or $R = I_{2i}/I_{2h}$, where indexes $h$ and $i$ denote the homogeneous and inhomogeneous magnetic field, accordingly. The degradation of the charged particle relative transverse momentum precision is proportional to $1 - R$.

In Figure 10, the magnetic field double integral degradation $1 - R$ with respect to homogeneous and inhomogeneous field integrals is shown for three cases: $1 - R_{mh}$, $1 - R_{bh}$, and $1 - R_{mb}$.

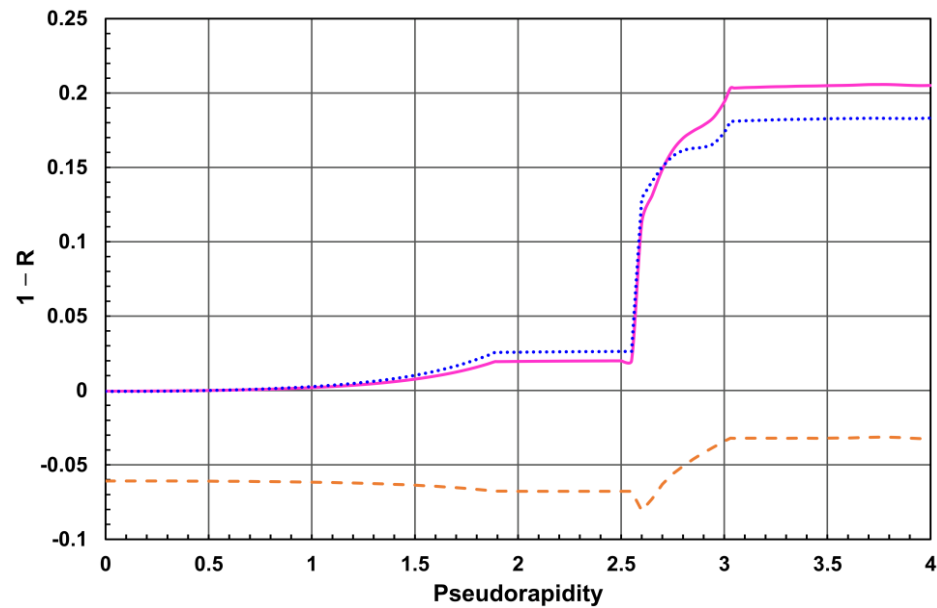

**Figure 10.** Magnetic field double integral degradation $1 - R_{mh}$ (solid line), $1 - R_{bh}$ (small-dotted line), and $1 - R_{mb}$ (dashed line) in the inner tracker of the FCC-hh detector vs. the pseudorapidity.

As shown in this Figure, in the pseudorapidity region from 0 to 1.88753 the degradation of the integrals rises as follows: $-0.0004 < 1 - R_{mh} < 0.0194$, $-0.0006 < 1 - R_{bh} < 0.0256$ and $-0.0607 > 1 - R_{mb} > -0.0676$.

In the pseudorapidity region from 1.88753 to 2.55, the degradation of the integrals varies slightly as follows: $0.0194 < 1 - R_{mh} < 0.0199$, $0.0256 < 1 - R_{bh} < 0.0263$ and $-0.0676 > 1 - R_{mb} > -0.0678$.

In the pseudorapidity region from 2.6 to 3.02982, the degradation of the integrals rises as follows: $0.1131 < 1 - R_{mh} < 0.2032$, $0.1294 < 1 - R_{bh} < 0.18$ and $-0.0808 < 1 - R_{mb} < -0.0320$.

In the pseudorapidity region from 3.02982 to 4, the degradation of the integrals varies slightly as follows: $0.2032 < 1 − R_{mh} < 0.2051$, $0.18 < 1 − R_{bh} < 0.1831$ and $−0.0320 > 1 − R_{mb} > −0.0324$.

These results finally show that in the transverse momentum resolution, the minimal steel yoke design exhibits an advantage in comparison with the baseline magnetic system between the level of 3.2–8.1%.

## 4. Discussion

In this section, we summarize the compared features of the baseline design [4] and the minimal steel yoke option of the magnetic system for the FCC-hh detector at the FCC.

The minimal steel yoke option, simulated with the coil parameters of the baseline magnetic system, offers the following advantages:

- 6% larger central magnetic flux density.
- 13.2% smaller compression axial force and pressure in the main coil middle plane.
- 16.2% smaller radial pressure in each forward coil.
- 12.1% shorter radial distance to the acceptable safety level of the magnetic stray field.
- 7.2% shorter axial distance to the acceptable safety level of the magnetic stray field.
- 3.2 to 8.1% larger magnetic field double integrals and, thus, better transverse momentum resolution of the charged particle registered in the inner tracker.

The minimal steel yoke magnetic system, simulated with the coil parameters of the baseline magnetic system, has the following disadvantages:

- 0.97% larger axial attractive force onto each forward coil.
- 9.3% lower magnetic flux density in the transition region between central and each forward coil.
- 5.8% larger stored energy in the magnetic system.
- 61.3% larger radial pressure in the main coil.
- The compression axial force to the central barrel wheel of 236.9 MN each side is 2.92 times larger than the similar axial force in the CMS magnet yoke at the 3.81 T central field.
- The attractive axial force to the yoke endcap assembly of 164 MN from each side is 2.42 times larger than the similar axial forces in the CMS magnet yoke at the 3.81 T central field.

However, the main disadvantage of the minimal steel yoke option would be the cost of the steel yoke construction which would negate all advantages of this option. The mass of steel used in the minimal steel yoke model amounts to 22,240 tons without considering the mass of the support structure. The cost of a 10,745 ton CMS yoke was estimated in 1997 [11] at CHF 43.8 million. Based on this estimate, the cost of the minimal steel yoke option could be at least CHF 90.7 million, which is more than twice the cost of manufacturing the CMS yoke.

Taking this cost into account, the relative gains from the minimal steel yoke design are not significant enough to warrant the additional expense. The iron-free baseline magnetic system [4] designed for the FCC-hh detector has almost the same electromagnetic features and similar charged particle momentum resolution as the simulated minimal steel yoke option.

## 5. Conclusions

This study investigates the advantages and disadvantages of a simulated magnetic system with a minimal steel yoke in comparison with the designed baseline iron-free magnetic system for the detector at the Future hadron-hadron Circular Collider with a center-of-mass energy of 100 TeV. Although the minimal steel yoke option offers certain advantages over the FCC-hh detector baseline magnetic system, the significant cost of the steel yoke construction would likely negate these advantages. The performed study confirms the choice in favor of the baseline design in relation to the simulated minimal steel yoke option.

**Author Contributions:** Conceptualization, V.K., A.B., H.G., A.H., M.M., W.R. and H.T.K.; methodology, V.K.; software, V.K. and M.M.; validation, V.K. and M.M.; electrical engineering, C.P.B., B.C. and A.D.; mechanical engineering, A.G., H.G. and U.W.; writing—original draft preparation, V.K.; writing—review and editing, V.K., H.G. and C.P.B. All authors have read and agreed to the published version of the manuscript.

**Funding:** This research received no external funding.

**Institutional Review Board Statement:** Not applicable.

**Informed Consent Statement:** Not applicable.

**Data Availability Statement:** No data are available.

**Acknowledgments:** The authors are very grateful to Claire Lee of Fermilab for much help with English corrections.

**Conflicts of Interest:** The authors declare no conflicts of interest.